# Sex-biased dispersal: a review of the theory

Xiang-Yi Li[1,*] and Hanna Kokko[1]

[1]*Department of Evolutionary Biology and Environmental Studies, University of Zurich, Winterthurerstrasse 190, CH-8057 Zurich, Switzerland*

[*]Author for correspondence: (E-mail: xiangyi.li@ieu.uzh.ch; Tel.: +41 (0)79 354 44 75).

ABSTRACT

Dispersal is ubiquitous throughout the tree of life: factors selecting for dispersal include kin competition, inbreeding avoidance and spatiotemporal variation in resources or habitat suitability. These factors differ in whether they promote male and female dispersal equally strongly, and often selection on dispersal of one sex depends on how much the other disperses. For example, for inbreeding avoidance it can be sufficient that one sex disperses away from the natal site. Attempts to understand sex-specific dispersal evolution have created a rich body of theoretical literature, which we review here. We highlight an interesting gap between empirical and theoretical literature. The former associates different patterns of sex-biased dispersal with mating systems, such as female-biased dispersal in monogamous birds and male-biased dispersal in polygynous mammals. The predominant explanation is traceable back to Greenwood's (1980) ideas of how successful philopatric or dispersing individuals are at gaining mates or the resources required to attract them. Theory, however, has developed surprisingly independently of these ideas: models typically track how immigration and emigration change relatedness patterns and alter competition for limiting resources. The limiting resources are often considered sexually distinct, with breeding sites and fertilizable females limiting reproductive success for females and males, respectively. We show that the link between mating system and sex-biased dispersal is far from resolved: there are studies showing that mating systems matter, but the oft-stated association between polygyny and male-biased dispersal is not a straightforward theoretical expectation. Here, an important understudied factor is the extent to which movement is interpretable as an extension of mate-searching (e.g. are matings possible *en route* or do they only happen after settling in new habitat – or can females perhaps move with stored sperm). We also point out other new directions for bridging the gap between empirical and theoretical studies: there is a need to build Greenwood's influential yet verbal explanation into formal models, which also includes the possibility that an individual benefits from mobility as it leads to fitness gains in more than one final breeding location (a possibility not present in models with a very rigid deme structure). The order of life-cycle events is likewise important, as this impacts whether a departing individual leaves behind important resources for its female or male kin, or perhaps both, in the case of partially overlapping resource use.

CONTENTS

# I. INTRODUCTION

Dispersal is defined as the movement of individuals or propagules with potential consequences for gene flow across space (Ronce, 2007; Saastamoinen *et al*., 2018). Dispersal exposes individuals to various costs, with possibilities including the energetic cost of movement (or that of traits that enhance passive transport, e.g. winged seeds), increased exposure to predators, failure to find a suitable site to settle in, and (in territorial species) hostile behaviours of resident individuals that aim to prevent immigrants from settling down (for a review see Bonte *et al.*, 2012). Despite these costs, dispersal is a ubiquitous phenomenon that occurs throughout the tree of life.

There are often biases in the propensity and/or distance of dispersal between the two sexes (Trochet *et al.*, 2016). In mammals, males often disperse more frequently and further away than females, the opposite pattern being typical for birds (Greenwood, 1980; Mabry *et al*., 2013; Trochet *et al*., 2016) where male-biased dispersal is common in only some groups (Anatidae in particular; see Clarke, Sæther & Røskaft, 1997). Fishes offer examples of male-biased (Hutchings & Gerber, 2002; Anseeuw *et al.*, 2008; Cano, Mäkinen & Merilä, 2008) as well as female-biased dispersal (Taylor *et al.*, 2003), as do reptiles (male-biased examples: Keogh, Webb & Shine, 2007; Dubey *et al.*, 2008; Ujvari, Dowton & Madsen, 2008; female-biased: Olsson & Shine, 2003). In insects, there are numerous cases of wing polymorphism (Andersen, 1997), sometimes with a dichotomous sex difference such that either males (Hamilton, 1979; Crespi, 1986; Godfray, 1988) or females are the wingless sex (e.g. Barbosa, Krischik & Lance, 1989; Wahlberg *et al.*, 2010; Hopkins *et al.*, 2015). In plants, pollen (that moves paternal genes) and seeds (that move both paternal and maternal genes) disperse at a different time and also have distinct dispersal ranges (Ghazoul, 2005). The review of Trochet *et al.* (2016) collected 257 species of vertebrates and arthropods for which sex-biased dispersal has been identified.

Empirical studies have identified a great variety of factors impacting dispersal, including environmental cues, development stage and physiological conditions of the organism, and the cognitive abilities of animals (Bowler & Benton, 2005; Nathan *et al.*, 2008; Morales *et al.*, 2010). Obviously, listing proximate factors alone is not enough to explain the ultimate causes behind dispersal evolution. It is often hard or not feasible to collect data and/or test hypotheses of dispersal in open populations (Ims & Yoccoz, 1997; Ruckelshaus, Hartway & Kareiva, 1997), especially when different driving forces of dispersal are often intertwined in a multicausal fashion (Starrfelt & Kokko, 2012*b*). The study of dispersal has therefore benefited greatly from explicit models that can clarify the steps of logic behind statements such as "even if habitats are always stable and dispersal is costly, individuals are still selected to move as long as this frees up resources to be used by related individuals" (a seminal finding by Hamilton & May, 1977). Similarly, Bengtsson (1978) showed that inbreeding avoidance can be a strong driver of dispersal. Thus, since the 1970s, mathematical models have been expanding our knowledge, and deepening our understanding of this complex problem.

There are excellent general reviews of the mechanisms and causes behind the evolution and maintenance of dispersal [see Ronce (2007) and Duputié & Massol (2013) for concise reviews, and Clobert *et al.* (2001, 2012) for book-length treatments]. Quite a large fraction of the theoretical dispersal literature, however, ignores sex differences in dispersal. Recent synthetic treatments of sex-biased dispersal (Dobson, 2013; Mabry *et al.*, 2013; Trochet *et al.*, 2016), on the other hand, focus on testing largely verbally expressed theories without providing a thorough review of the relevant developments of mathematical models in the field.

Therefore, we assess here the theory of dispersal through the specific lens of whether it predicts sexes to differ with respect to this trait. Understanding dispersal, to the extent that it can be viewed as a trait undergoing adaptive evolution (i.e. excluding accidental or incidental gene flow, see Burgess *et al.*, 2016), requires specifying the fitness consequences that arise through moving. In one sense, it is then easy to understand that sex differences can arise. If, say, sexual selection causes body-size differences, and the physical costs of moving are body size dependent, then dispersal costs will not be identical for the two sexes – and one has identified a potential asymmetry that might explain why one sex is more sedentary than the other (Gros, Hovestadt & Poethke, 2008). As we will see, however, theoretical predictions rarely boil down to simple statements that the costs differ by a certain magnitude while the benefits are identical across the sexes, or *vice versa*. The accumulated body of theory is more multifaceted than this because (1) the fitness effects of dispersal can be direct or indirect, and (2) selection on sex-biased dispersal can also show coevolutionary patterns, i.e. selection operating on one sex cannot be understood without considering how often, and how far, individuals of the other sex have evolved to disperse.

For these reasons, we aim to do more than simply listing all possible causes of asymmetric fitness consequences of residing in the natal habitat *versus* moving elsewhere. Instead, our aim is to provide an overview of arguments (and their interactions) based on different fitness consequences of dispersal; for each of the potential drivers identified by theory to date, we ask how it might operate differently for the two sexes. We consider (1) asymmetric limiting resources (including mating opportunities) and the competitive ability of dispersers *versus* non-dispersers (classic reference: Greenwood, 1980), (2) kin competition (classic reference: Hamilton & May, 1977), (3) inbreeding avoidance (classic reference: Bengtsson, 1978), and end with more complex settings where (4) stochasticity and genetic architecture matter. Each of the drivers can, at least potentially, differ between the sexes, in ways we outline below; note, however, that the classic references cited above differ greatly in how much they focused on sex differences or on explaining why any organism, regardless of sex, should disperse in the first place [emphasis on sex differences is the main theme of Greenwood (1980), but a minor point in Hamilton & May (1977)]. It is also important to note that these driving forces are not mutually exclusive alternatives as determinants of the propensity and distance distribution of dispersal of each sex. Instead, they almost always interact in nature (Starrfelt & Kokko, 2012*b*), and in theoretical work, it has become common practice to consider several causal routes simultaneously.

The amount of attention paid to the different driving forces appears to differ between theoretical and empirical work. General theories of dispersal tend to lean rather heavily on the theoretical developments that have their origins in papers by Hamilton & May (1977) and Bengtsson (1978), who introduced simple models highlighting that kin competition and inbreeding, respectively, can select for dispersal even if performing it is costly. Empirical studies, on the other hand, are much more likely to choose Greenwood (1980) as their classic reference. Greenwood's (1980) paper outlined differences in determinants of success (when philopatric or dispersing) in male and female birds and mammals, but is cited widely beyond these taxa too. To use the first half of 2018 as a representative example (date of investigation of citation data: June 30), Hamilton & May (1977) had been cited 12 times in these six months, and only four of those papers (33%) were of empirical nature (if we allow both original data and literature reviews or meta-analyses of published data to count as 'empirical'). By contrast, Greenwood (1980)'s 44 citations during this period almost solely (39 papers, 89%) arose through empirical work using the criteria above, increasing to 40 (91%) if we include Andersson (2018) who discusses issues arising in two specific taxa (New World quails and waterfowl). We return to this issue in Section III.

**II. DRIVERS OF SEX-BIASED DISPERSAL**

**(1) Asymmetric limiting resources and the competitive success of philopatric and dispersing individuals**

In many environmental and social settings, the resource that limits fitness differs between males and females. Greenwood's (1980, 1983) verbal accounts are based on this insight, combined with knowledge of across-species patterns. These early papers pointed out that (socially) monogamous birds often have female-biased dispersal, whereas mammals often exhibit male-biased dispersal. Recent reviews have confirmed this pattern for these taxa (Mabry *et al.*, 2013; Trochet *et al.*, 2016), although exceptions exist (reviewed in Lawson Handley & Perrin, 2007). Against this background, it is interesting to note that Greenwood's arguments differ rather substantially from those assumed by most theoretical models. The key difference is that Greenwood focuses on *direct* success (mate and/or resource acquisition) of philopatric *versus* dispersing individuals, while theory is more often based on the effects that dispersal imposes on spatial relatedness patterns, together with the indirect fitness effects that arise because competition is relaxed at the natal deme when the focal individual no longer competes for resources there (see Sections II.2 and II.3). These latter effects can exist even if every individual – whether it has dispersed or not – is an equivalently strong competitor, and we show below that this is sufficient to create a rich set of alternative patterns of sex-biased dispersal.

We suspect that a reason why models generally do not begin with 'Greenwoodian' ideas is that his work jumps straight into understanding sex differences, without focusing on why to perform (costly) dispersal in the first place. Formal models do not work unless they incorporate a mechanism favouring dispersal, and they usually evoke kin selection, inbreeding avoidance or spatiotemporal stochasticity to achieve this (Starrfelt & Kokko, 2012*b*). Still, theoreticians should not remain blind to the idea that direct fitness consequences of mobility can be sex specific. To incorporate 'Greenwoodian' mechanisms, a model should specify direct fitness consequences that apply to the disperser itself: how much easier it is for a philopatric individual to gain access to a resource compared to an immigrant. Rather surprisingly, we are aware of only one study (Perrin & Mazalov 1999) where this is done explicitly with designated parameters capturing the intended effects: $a_M$ and $a_F$ denote the (possibly lower) competitive value of an immigrant male or female relative to that of a philopatric individual of the same sex. The lower competitive success of dispersers corresponds to a sex-specific dispersal cost, and this cost is expected to be higher (with a lower value of *a*) for the sex that is responsible for territory acquisition. The model combines the potential asymmetry ($a_M \neq a_F$) with inbreeding avoidance, and as we will explain in more detail in Section II.3, the prediction is that one sex is often predicted to carry all the burden of costly dispersal (together with phylogenetic constraints: sexes may evolve to be stuck in a sex-biased setting even if the dispersal cost structure changes later; see Perrin & Mazalov, 1999).

It appears to us that there is scope for substantial work evaluating how parameters such as $a_M$ and $a_F$ relate to real mating systems. Perrin & Mazalov (1999) rephrase Greenwood's argument – that it is easier for a male bird to establish a territory at, or near, his natal site than to achieve the same elsewhere – as males 'taking the responsibility for' acquiring the pair's territory ($a_M < a_F$). Consequently, when turning their attention to mammals, they argue that $a_M > a_F$ is possible whenever females take this responsibility. This implies, however, that female mammals are strongly penalized (small $a_F$) if they emigrate, while male success is less strongly determined by whether he is already familiar with an area or not. This appears, to us, a surprising poor fit to typical cases of mammalian polygyny, that often feature stronger inequalities of male rather than female mating success (harem ownership is a difficult achievement for males, while females are rarely prevented from entering harems and using the local resources). Greenwood's original argument for polygynous, dispersive males (as found in a typical mammal) shows a similar lack of clarity compared with his argument for birds. While one might accept that the mammalian need to defend females rather than local sites may lessen the importance of any home advantage for males, his work does not articulate very clearly why male-biased (rather than unbiased) dispersal becomes the norm, unless one assumes similarly strong familiarity-based competition arguments to apply to mammalian females as to monogamous territorial male birds.

It is, obviously, not an easy task to measure how competitive a male (or a female) *would* have been *if* it had done the opposite from what it did (tried to stay if it dispersed in reality, or *vice versa*); however, such comparisons – perhaps experimentally achievable in some settings – are required before one can say much about the values of $a_M$ and $a_F$ in real settings. In the model of Perrin & Mazalov (1999), the values of $a_M$ and $a_F$ never exceeded 1, i.e. emigrating never improved an individual's competitive ability (dispersal was always costly). Immigrants might, however, also enjoy elevated success if choosy females prefer them as mates (Motro, 1991; Lehmann & Perrin, 2003), which in the current context would imply $a_M > 1$. Future studies investigating the entire range of options could also combine such effects with sex-specific dispersal mortality (Wild & Taylor, 2004; Gros *et al.*, 2008). While there is at first sight no mathematical reason to differentiate between a low value of *a* (low competitive ability of an immigrant) and high dispersal mortality, the population dynamic consequences may show interesting feedback if, for example, one sex needs to travel further before it can settle, perhaps because of low mortality in one sex causing biased adult sex ratios that in turn make the (territorial) world more crowded from the perspective of one of the sexes [for a unisexual case see McCarthy (1999)]. There is clearly scope for new models to help understand how Greenwood's idea translates to mathematically derived expectations.

There is also a section of Greenwood (1980) that has gained little subsequent attention, and which likewise would be interpreted as an '*a* > 1' case, in the sense of dispersal leading to direct improvements of fitness. He extrapolated an argument that is based on mate-searching: males should move more whenever this helps them to find more fertilization opportunities (this obviously requires polygyny), and this extra movement ultimately leads to a pattern with male-biased dispersal. Later in the paper he then turns to species-specific mechanisms of mate acquisition and defence, elucidating when a male bias may or may not be realized.

This idea is interesting, especially because theoretical studies often show no bias (including the case where the direction of the sex bias is random) or female-biased dispersal in polygynous settings where one might expect male-biased dispersal to evolve easily (Perrin & Mazalov, 1999, 2000; Guillaume & Perrin, 2009; Hovestadt, Mitesser & Poethke, 2014; Henry, Coulon & Travis, 2016). Studies examining the effect of mating systems on dispersal often contrast competition over mates with competition for other resources, phrased as local mate competition (LMC) (Hamilton, 1967) and local resource competition (LRC) (Clark, 1978), which are typically thought to be a problem for males and females, respectively. At first sight, the matter appears clear: male-biased dispersal evolves under polygyny if the intensity of mate competition between males exceeds all other types of competition, including the competition between females for food and territories (Dobson, 1982; Lawson Handley & Perrin, 2007; Brom *et al.*, 2016). However, the ease with which models actually generate the condition LMC > LRC (in terms of strength of competition) is less straightforward than intuition might suggest.

Consider a common model formulation, where female reproductive output is limited by the number of sites in a local patch (also called deme or site; e.g. $N$ females can breed per deme). Also assume that there are equally many females as there are males in the population as a whole. Then local males compete over precisely $N$ reproductive opportunities, and females compete over $N$ sites. Even though fewer males might succeed in using these $N$ opportunities than the $N$ females who each succeeds in securing one breeding opportunity, the mean success is identical across the sexes; in other words, LMC is equally strong as LRC (Perrin & Mazalov, 2000). Consequently, polygyny and its 'intuitive' difficulties of mate acquisition for males do not automatically predict males to become the dispersing sex (Perrin & Mazalov, 2000). While variance in breeding success will be sex specific, this does not translate straightforwardly to strong sex biases in dispersal either (see Section II.4 for further details).

It is possible to remove all competition between females (e.g. the second model of Perrin & Goudet, 2001) which then establishes male-biased dispersal and female philopatry, but also results in open-ended population growth. While conceptually illuminating, this is only realistic if population regulation is assumed to operate at some other time of the year [as explained by Perrin & Mazalov (2000) in their model variant where females are not limited by resources]. Ideally, a theoretical study should not merely mention this problem, but would also explicitly model the spatial scale of this regulation, for if any regulation operates locally, there is still the chance that a departing individual leaves behind resources that are necessary for survival, and thus improves the chances that its kin will reproduce later (and kin selection is an important driver of dispersal, see Section II.2).

In this context, more explicit modelling of Greenwood's original idea of male mate-searching could prove illuminating. The phenomenon of the 'identical $N$' arises in models that assume a strict deme structure, which is a spatial assumption that does not necessarily capture all aspects of mate and resource competition in nature. Specifically, any model that assumes all matings of an individual to occur in one patch (or deme) automatically excludes any male mobility benefit that is based on exploiting fertilization opportunities in more than one patch (Greenwood, 1980).

We therefore now turn to models that consider more flexible settings, and avoid the assumption of discrete steps from dispersal to mating (in one patch) followed by reproduction. For example, Hirota (2005) showed that female-biased dispersal can evolve even if they mate with multiple males, as long as female remating is not synchronized with dispersal or they do not only use the sperm from their last mate. Other models allow individuals to assess their current prospects using indirect cues such as local density of competitors or opposite-sex individuals. Once emigration decisions are allowed to depend on local densities of males and/or females (Hovestadt *et al.*, 2014), male-biased dispersal becomes easier to explain than in the classic models, while female-biased exceptions also become explicable under special circumstances.

Shaw & Kokko (2014) additionally consider that individuals on the move do not necessarily follow a predefined dispersal kernel that forces them to land and settle irrespective of local conditions. Animals with sufficient cognitive capacities might instead evaluate each habitat patch for key fitness indicators, a very relevant indicator being the number of males and females already present in each patch. If individuals continue searching for new patches as long as the most recently encountered sex ratio is an unfavourable one (e.g. for males, a male-biased one), but accept to settle probabilistically in non-ideal settings too (as each step moved has costs), the outcome turns out to depend not only on monogamy *versus* polygyny, but also on whether mating happens before, during, or after dispersal.

Such work (Shaw & Kokko, 2014) also provides a potential resolution to an apparent contradiction when two relatively disconnected sets of literature – that of mate searching and that of dispersal – are considered jointly. If one sex is highly mobile in its mate-searching efforts, then, in the mate-searching literature, it is often found that the other sex can save the effort and move very little or not at all, as it 'will be found' in any case (Hammerstein & Parker, 1987; Shaw & Kokko, 2014; Fromhage, Jennions & Kokko, 2016). But if sex-specific dispersal is allowed to evolve in a setting where densities and local sex are not artificially kept constant for analytical purposes, the opposite is found: elevating female dispersal will also make males disperse more (Meier, Starrfelt & Kokko, 2011). Philopatry now becomes a poor option when individuals of the other sex often disperse, because the philopatric individuals are above average often born in patches that just had a high breeding success in the previous generation, and this makes them lose more opposite-sex emigrants than they on average receive, *via* immigration from other patches, as potential mates. This asymmetry is an automatic consequence of demography itself (non-dispersers happen to be particularly numerous in the sites that have above-average productivity in the previous generation) and does not depend on any particular dispersal rule, but if the model ignored temporal variation in breeding success, the effect would vanish. In Shaw & Kokko (2014), the dispersal of the two sexes can be either positively or negatively correlated, and the details depend on the extent to which female movement is interpretable as mate-searching (i.e. the timing of mating relative to movement). They recover the prediction that it is difficult to find a sex bias in dispersal if mating and reproduction occur after dispersal in the final settlement patch, even if matings are promiscuous; mating *en route* on the other hand favours male-biased dispersal (with details that depend on first- or last-male sperm precedence), while pre-dispersal mating favours female-biased dispersal.

The above highlights that it can be important to understand mobility in the context of variable population densities, Allee effects (low reproduction at low densities, potentially because of mate-finding difficulties) and individual decision-making rules (Gilroy & Lockwood, 2012). Mobile individuals can increase encounter rates with others beyond simple deme-structure assumptions, thus future models could consider the evolving sex specificity of scales over which competition for resources or mates occurs [for a spectacular empirical example, consider male pectoral sandpipers (*Calidris melanotos*) that can move through a considerable part of the species' breeding range, up to 13,000 km, during one mating season (Kempenaers & Valcu, 2017)].

To summarize, the fitness of males and that of females are often limited by different types of resources (including the importance of mate availability). When the familiarity with local environments benefits one sex more than the other in competition with same-sex individuals, sex-biased dispersal appears intuitively obvious, but rigorous empirical tests appear equally scarce as formal models of the idea. Polygyny *per se* does not necessarily lead to male-biased dispersal in models where local resources limit female reproduction. The spatial scales over which competition occurs (and over which reproductive success can be gathered) matter, but remain often unquantified. When the spatial structure is not limited to separate island-like demes and mating and reproduction can happen multiple times in a lifetime, dispersal can increase the rate of encounter with individuals of the opposite sex. There is much scope for empiricists to test assumptions (not merely predictions) of verbal and mathematical models, and for theoreticians to incorporate ideas that reflect biological diversity in the order of life-cycle events over varying spatial scales.

**(2) Kin selection**

(*a*) *Kin competition within the same generation*

Above, we commented on local mate competition and local resource competition as potential drivers of dispersal, but have not yet explained why individuals are selected to undertake risky journeys in the first place. If competitors exist elsewhere too, it is not easy to understand why emigration can pay off (unless the natal habitat is ephemeral, or if environmental conditions fluctuate drastically, leading to suitable and unsaturated habitats elsewhere). The work of Hamilton & May (1977) solved this puzzle by identifying a strong effect on *indirect* fitness. Kin competition can select for dispersal even if the environment is static, the population is saturated, and all individuals are equally strong competitors no matter where they reside.

The first model of Hamilton & May (1977) does not consider sexual reproduction. While it is thus not a model suitable for making predictions about sex bias, understanding it is important for any subsequent discussion of the inclusive fitness effects that can arise based on an individual's presence or absence at a site. The authors assume a population of a parthenogenetic species living in an environment with a fixed number of sites. In each site only one adult can survive. At the end of each year, all adults produce the same number of offspring and die. If there is no dispersal, all offspring produced on the same site compete with each other, leaving only a single survivor. A rare mutant strategy, that allows only one of the offspring (of the mutant) to stay and makes the rest disperse randomly to other sites – where they compete on equal terms with the offspring there – is bound to have higher fitness. This is because the single offspring that stays in the natal site has no competition and thus will certainly survive and reproduce, while the dispersers each have a chance to colonize other sites. In this way, the mutant dispersal strategy is likely to spread, and the consequent competition by immigrants makes fitness of different strategies frequency dependent, with an evolutionarily stable outcome of non-zero dispersal probability.

The first model in Hamilton & May (1977) shows clearly that kin competition can be a powerful driver of dispersal in asexual populations. Being asexual, their basic model obviously did not comment on sex differences, but their paper also includes one modelling extension that allowed the authors to discuss some consequences of sexual reproduction. They note that sexuality comes with the possibility of inbreeding, which will complicate the computations of fitness expectations. To consider a case of minimal complexity, the authors assumed that all males always disperse, and ask whether females should also disperse or not, if their subsequent breeding follows the same rules of competition as before. The authors could then show that sexual reproduction lowers the evolutionarily stable dispersal probability of females (Fig. 1), as parent–offspring relatedness is lower than under asexuality.

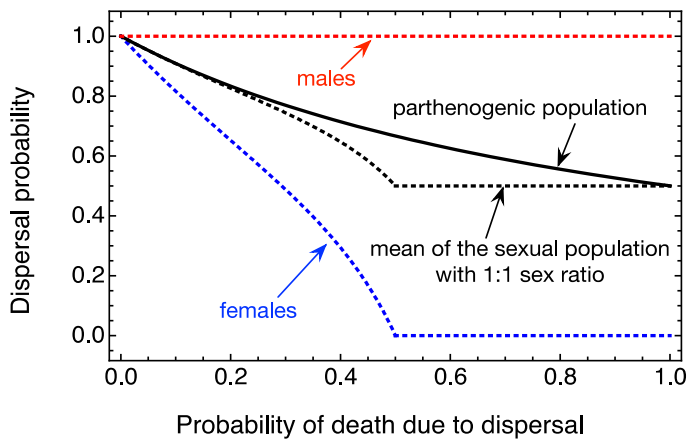

Figure 1. Evolutionarily stable (ES) dispersal probability as a function of the mortality rate in the model of Hamilton & May (1977) in asexual and sexual populations.

Kin competition is so ubiquitous in mathematical models of sex-biased dispersal that making it completely ineffectual is almost always intentional. In Section II.1 we already discussed how ideas of LMC and LRC relate to Greenwood's ideas, but as Greenwood's focus was on direct fitness effects of dispersing, it is important to be reminded that kin competition is the key ingredient of models where LMC and/or LRC drive dispersal. There are techniques to remove the effect of kin competition, which can be used to test its importance compared with other factors. For example, Perrin & Mazalov (1999, 2000) created conditions where kin competition cannot operate by artificially setting the relatedness within patches to zero, in order to isolate the effect of inbreeding avoidance. Similarly, in the model of Hovestadt *et al.* (2014), kin competition is removed from the model by assuming unlimited resources and by letting the population grow geometrically (all mated females are expected to produce the same number of daughters independent of the population density in their patches), in order to isolate the effect of fluctuating local sex ratio on the evolution of male dispersal probability. The book chapter of Perrin & Goudet (2001) used a similar approach of eliminating resource competition (in their section *Inbreeding without competition*) for isolating the effect of inbreeding depression on the evolution of sex-biased dispersal.

In individual-based models of dispersal, kin competition naturally emerges, but here too its effect can be prevented on purpose. For example, to confirm that kin competition indeed is the driver of sex-biased dispersal, Brom *et al.* (2016) created a case where kin competition is deliberately eliminated by a shuffling procedure that destroys relatedness structure while leaving the demographic structure and ecological settings unchanged [similar to Poethke, Pfenning & Hovestadt (2007) where dispersal was not sex-specific]; this made the sex bias in dispersal disappear. The strength of the effects found suggests that the empirical focus on 'Greenwoodian' direct fitness effects could be very usefully complemented by attempts to quantify the competitive effects caused by the presence of kin. Assessing (or experimentally manipulating) local mate and resource competition is not impossible (e.g. Shuker *et al.*, 2005; Silk & Brown, 2008), and there is clear scope for conducting more studies that examine how these factors relate to sex-biased dispersal (Baines, Ferzoco & McCauley, 2017).

Empirical quantification difficulties might not be the only reason why kin competition has been more popular with theoreticians (with their idealized demes) than with empiricists (who often place their study areas in continuous, large populations). It may be difficult to perceive kin competition as a strong evolutionary force whenever considering real demes with their fluid, difficult-to-define boundaries. The degree of population subdivision, however, is itself a function of dispersal, and this leads to an interesting conceptual issue. Consider a hypothetical ancestral population with limited dispersal and small demes, where kin competition is consequently strong. These conditions select for dispersal, and the area over which individuals move and settle evolves to be much larger. A snapshot of this later situation might tempt an interpretation that kin competition is not a problem because individuals disperse far enough to avoid it – in other words, evolution has solved a problem to the extent that the original driver of the trait is barely perceptible. This problem is not restricted to dispersal evolution; it is a general one in evolutionary ecology, a classic example being the 'ghost of competition past', i.e. species that do not currently appear to compete over the same niche space may have evolved to avoid doing so (Connell, 1980).

Turning back to sex specificity in kin contexts, an understudied question remains: the stay-at-home relative's success is not necessarily increased equally much when the dispersing individual is a male or a female, and the sex of the staying relative is obviously of importance too. Models such as Perrin & Mazalov (2000) and Brom *et al.* (2016) address maximally contrasting cases in terms of the mating system: within each paper, monogamy is included to make the fitness determinants of the two sexes maximally identical, and the results are contrasted with various non-monogamous cases, where males are assumed to compete for fertilization opportunities (and nothing else), and only females are interested in resources other than fertilizations (e.g. nest sites, food).

While stark contrasts are conceptually useful, the extreme sexual segregation of the assumed competition patterns also creates a knowledge gap. If we assume, like Perrin & Mazalov (2000) did, that a male's presence on a patch does not deplete the food or have any other ecological effect on the success of local females (that may include his sisters), we cannot say much about the expected sex bias should there be some intersexual competition for survival-related resources as well. Brom *et al.* (2016) use a different structure, where density dependence impacts the survival of all individuals. Although at first sight this creates intersexual competition as brothers' presence does harm their sisters, the density dependence is applied after birth and before dispersal (generations are non-overlapping). This means that there is no scope for dispersal to alter the way the two sexes can survive in the presence or absence of each other. There are interesting but understudied complications because, typically at least, both sexes need nutrition that they acquire locally, and other non-shareable local resources may be similarly important for survival (in some species, space itself is a limiting resource). On the other hand, males compete among each other for fertilization opportunities in a way that does not occur between the sexes. In some situations, females may also compete for access to males. The potential for these variations to impact sex-specific dispersal appear understudied.

The study of Brom *et al.* (2016) highlights a generally important point: the importance of kin competition can be adjusted by changing the order of life-cycle events. If most sibling competition is scheduled to happen after a time window for dispersal, then elevating the dispersal rate can relax competition among siblings, but the same effect cannot occur if competition already happens before dispersal is possible. Investigating effects of the order of events in the life cycle has become commonplace in general models of dispersal (e.g. Bach *et al.*, 2006; Massol & Débarre, 2015), but sex-biased dispersal models have so far seldom made systematic comparisons across life cycles with different orders of life-cycle events (we return to this in Section II.2*b*), or, more generally, population regulation within and across the two sexes. One recent study (Henry *et al.*, 2016) is a welcome exception: the authors point out that if we assume that local sex ratios will be kept more even – in their study it is kept completely constant – under sex-specific density regulation (male presence does not impact female survival or *vice versa*) than under joint regulation (survival of all individuals is impacted by all others), then joint regulation may increase male-biased dispersal, for reasons of altered between-patch fitness variances (we return to this study in Section II.4).

(*b*) *A better studied corner: coevolution between sex-biased dispersal and primary sex ratio*

Interestingly, the question of sex-specific kin competition appears most intensely studied in the very specific context where it coevolves with sex-ratio biases (Leturque & Rousset, 2003; Wild & Taylor, 2004). Here, researchers have also explored the effect of the order of life-cycle events more thoroughly than in the case where dispersal is the only evolving trait. The key reason why dispersal and sex ratio might coevolve is that the intensity of competition (for a resource that only one of the sexes is interested in) depends both on the primary sex ratio (Werren, 1980; Herre, 1985; Macke, Olivieri & Magalhães, 2014) and on dispersal (with its potential to be sex-biased; Leturque & Rousset, 2003). One can then study how an existing sex bias in dispersal impacts sex ratio (Pen, 2006; Guillon & Bottein, 2011), or ask an even more general question: what is the expected coevolutionary pattern when both sex ratio and sex-biased dispersal can evolve?

If males compete with each other for access to females, from their point of view, the patch-specific sex ratio determines the quality of the patch in terms of expected reproductive success. Therefore, a male-biased sex ratio can promote male dispersal analogously to low-quality habitat promoting emigration in general (Taylor, 1988; Nelson & Greeff, 2011). If the sizes of habitat patches are similar to each other, the sex ratio is also an indication of the number of reproductive females, which determines the relatedness among the offspring and the intensity of kin competition (Nelson & Greeff, 2011). But the net outcome of all these effects depends on the order of life-cycle events, which impacts the composition of the set of individuals interacting with each other when competition over limited resources occurs (Wild & Taylor, 2004). Specifically, if dispersal occurs before mating, sex ratios are expected to evolve a bias towards the sex that is less likely to compete in its natal patch (Bulmer &

Taylor, 1980). Here it makes sense to produce many individuals of the sex that, due to dispersal, avoids much of the competition with its siblings. This situation becomes more complicated if males disperse before mating whereas females disperse after mating (carrying sperm with them): female-biased sex ratios evolve when males are more likely to compete locally, but the reverse is not always true; male-biased sex ratios are not as easily achieved by this model (Wild & Taylor, 2004), which reflects its (justifiably) asymmetric assumption set – situations where males take the fertilized eggs with them after mating are not common [but not logically impossible: for tadpole-carrying male frogs, see Ringler *et al*. (2016) and for egg-carrying male bugs, see Reguera & Gomendio (1999)].

Wild & Taylor (2004) also produced a somewhat counterintuitive result: despite coevolution, knowledge of primary sex ratio is not required for predicting evolutionarily stable dispersal rates. Nelson & Greeff (2011) point out that this result ceases to be true if one includes environmental and/or demographic stochasticity, as in this case the sex ratios no longer remain homogeneous across habitats. Without heterogeneity in sex ratios, a male cannot expect to improve his chance of finding a patch with a more favourable sex ratio by dispersing. Using an individual-based modelling approach with demographic stochasticity, and assuming that different genes determine dispersal probabilities from patches of specific sex ratios (this implicitly assumes that individuals can measure the local sex ratio and use the information for making dispersal decisions), Nelson & Greeff (2011) show that the probability of dispersal can evolve to be dependent on the sex ratio.

(*c*) *Kin competition and parent-offspring conflict*

In species with overlapping generations, kin interactions are not restricted to occur within a generation. Conflict of interest between parents and offspring (Trivers, 1974; Motro, 1983,

Wolf & Wade, 2001) may form a powerful driving force for sex-biased dispersal, especially if competition solely, or mostly, occurs within one sex. Here, the control of the dispersal phenotype matters: given the observation 'an individual has moved', is this best understood as the individual's gene being expressed – or perhaps those of its parent(s)? One example of the latter is prenatal exposure to corticosterone in the viviparous common lizard *Lacerta vivipara*: longer exposure increases juvenile dispersal in accordance with expectations based on mother–daughter competition (Vercken *et al.*, 2007). In plants, the seed coat together with any dispersal structure, and fruits (that may be attractive to frugivores who then disperse seeds) express the mother's genotype, rather than any offspring genes.

Offspring dispersal rates are expected to be higher if determined by the mother rather than by the offspring themselves (Motro, 1983). This also holds if the evolving trait is dispersal distance (Starrfelt & Kokko, 2010). It is in the mother's interest to 'manage' risk and make offspring go far enough to maximize her total reproductive success. The perspective of an individual offspring differs from this, as each would benefit by making their siblings take the high-risk option of high mobility, while themselves settling at a distance that is easier to survive. Starrfelt & Kokko (2010) noted that considering sex specificity of dispersal in the context of parent–offspring conflict is an obvious next step. Interestingly, there are old verbal predictions in this realm [Liberg & von Schantz, 1985; see also Marks & Redmond (1987) and Liberg & von Schantz (1988)], but these appear not to have caught theoreticians' attention so far. This set of papers discusses the idea that sex-biased dispersal may result from parents benefitting from avoiding a prolonged interaction with offspring of one particular sex. The specific question is whether philopatric daughters can cause fitness loss to their parents through nest parasitism. As this appears possible in birds but not in mammals, there might be selection for avian parents specifically to evict female offspring. Although newer literature has highlighted that intraspecific parasitism involving kin is not necessarily detrimental to the parent (Andersson, 2017), it appears important to consider the extent to which mothers, fathers, sons and daughters compete for the same or different resources.

This angle could help link the sex-biased dispersal literature with that on sex-specific conflicts regarding the right to reside in an area. Often these conflicts arise in social species (Ekman & Sklepkovych, 1994; Cant *et al.*, 2010), where a key question is whether the presence of philopatric individuals is helpful or harmful for their relatives (here harm may refer to active behaviours or simply depletion of shared local resources). The philopatric sex often evolves to be more helpful (or less harmful) than the dispersive sex, especially if there is generational overlap (Johnstone & Cant, 2008). Sex-biased dispersal can even help solve a general problem in social evolution theory regarding so-called *viscous populations* (i.e. those with limited dispersal leading to spatial relatedness structure): while it can be beneficial to show altruistic behaviour in the natal patch as there is positive relatedness between donors and recipients, the same population structure elevates competition among kin, which has the potential to cancel out any beneficial effects (Taylor, 1992). If, however, gene flow operates *via* both sexes, and one sex disperses less than the other, competition can be eased without completely destroying the relatedness structure (Gardner, 2010). There is clear scope for more work because the models tend either to assume an *a priori* existing sex-biased dispersal pattern and derive the consequences for social behaviour (Johnstone & Cant, 2008; Gardner, 2010; Johnstone, Cant & Field, 2012; Kuijper & Johnstone, 2017), or to ask coevolutionary questions about dispersal and helping, but without sex specificity (Mullon, Keller & Lehmann, 2018).

Parent–offspring conflict is also relevant for models that link sex-ratio evolution with sex-biased dispersal (Pen, 2006). If the offspring sex ratio is under maternal control, and the offspring of one sex is more expensive to produce or rear than the other, the sex ratio should be biased towards the 'cheaper' sex (often males) (Trivers, 1974), and the dispersal rate of the 'cheaper' sex should also be higher (Taylor, 1988; Gandon, 1999). The mother's reproductive success is optimized if just enough sons remain in the natal patch to fertilize the daughters, whereas the rest disperse to reduce local competition and also to maximize the chance of spreading her genes to other habitats. From the viewpoint of her sons, however, dispersal may be costly (e.g. through mortality) with no guarantee of mating success elsewhere. Therefore, if sons can control their own dispersal, the dispersal probability should evolve to be lower than the rate favoured by the mother; if the mother then still has the power to control the sex ratio, fewer males may be the result.

(*d*) *Summary of the role of kin competition in driving sex-biased dispersal*

To summarize, kin competition is a fundamental driving force for dispersal. It almost always plays a role in theoretical models of sex-biased dispersal, while empirical efforts to estimate its effect appear rarer, perhaps because it is easier to study kin interactions among individuals who are still jointly present than to quantify causal relationships from an individual's departure to the elevated success of the remaining individuals. Remarkably, sex-specific kin competition (and its dependence on life-cycle order effects) has been studied most intensively in the context where it coevolves with primary sex ratio. Since kin competition does not only operate within the same generation, there is also ample scope for parent–offspring conflict to drive dispersal patterns. While several effects of within- and across-generation kin competition on sex-biased dispersal have been studied, competitive interactions are often assumed to be constrained to operate within each sex, while intersexual competition is rarely taken into consideration.

**(3) Inbreeding avoidance**

Inbreeding, the production of offspring from matings between genetically related individuals, can increase the chance of homozygosity of recessive, deleterious alleles in the offspring (Charlesworth & Charlesworth, 1999; Szulkin *et al.*, 2013). The possible detrimental effects are captured under the umbrella term of inbreeding depression, and can include reduced birth mass, survival, reproduction and resistance to diseases in mammals and birds, decreased seed set, germination, survival and resistance to stress in plants (Keller & Waller, 2002), in addition to various taxon-specific effects in other taxa (e.g. development time in various invertebrates; DeRose & Roff, 1999). Because individuals that are far away from each other are unlikely to mate, inbreeding depression can select for disperal (Bengtsson, 1978; Waser, Austad & Keane, 1986; Perrin & Mazalov, 1999, 2000; Guillaume & Perrin, 2009; Hardouin *et al.*, 2015; Henry *et al.*, 2016), although the astute reader will note that this statement makes implicit assumptions about the timing of mating – for dispersal to help, it should precede mating.

Inbreeding avoidance is an intricate problem because of three interesting insights. First, the production of inbred young does not necessarily impact the fitness of males and females equally; even the signs of the fitness change can differ between the sexes, because the opportunity costs of each mating can be strongly sex specific. At the extreme, a female might fertilize her only egg of the whole breeding season with the sperm of a close kin, leading to low survival prospects for the young, while for the sire, the same mating might represent an additional opportunity that only adds to his success (even if only moderately, due to the young being inbred) if his chances with other females remain unchanged [for a detailed discussion of sex differences in inbreeding tolerance, see Waser *et al.* (1986), Kokko & Ots (2006), Lehtonen & Kokko (2015) and Duthie & Reid (2016) in a non-dispersal context, and Perrin & Mazalov (2000) in a dispersal context]. Second, if one sex routinely disperses far away from the natal site, the inbreeding problem has also been 'solved' for the other sex. Third, inbreeding avoidance will not automatically evolve as soon as there is inbreeding depression, because inbreeding also brings about a numerical benefit of transmitting alleles to the next generation (Fisher, 1941; Kokko & Ots, 2006; Duthie, Bocedi & Reid, 2016).

In the majority of models considering inbreeding and dispersal, inbreeding avoidance interacts with other forces to determine the probability of dispersal, including kin competition, mortality cost of dispersal, mating systems, and environmental and/or demographic stochasticity. To understand how the three insights can lead to sex-specific predictions, it is useful first to look at a model that considers inbreeding avoidance as the sole reason to disperse: the island population model of Perrin & Mazalov (1999).

In Perrin & Mazalov (1999), there are infinite numbers of demes, each offering an equivalent number of breeding sites, which in a saturated habitat equals the number of breeding females. The evolving traits are the sex-specific dispersal probabilities $m_M$ and $m_F$ for males and females, respectively. The mortality cost due to dispersal (expressed as survival $s < 1$) is the same for both sexes, and there is an additional cost of reduced competitiveness of immigrants, which may be sex specific: male and female immigrants are less competitive than natives, by a factor of $a_M$ and $a_F$, respectively, in obtaining breeding opportunities (see Section II.1). To account for the cost of inbreeding, the authors assume that female fecundity decreases linearly with coancestry between mates. The authors then calculate equilibrium coancestry

(which decreases with deme size), the probabilities of staying in the natal deme for males, $k_M$, and females, $k_F$, and the cost of inbreeding, $i$, and relate these to the cost of dispersal for males ($c_M = 1 - sa_M$) and females ($c_F = 1 - sa_F$). By balancing the trade-off between the benefits (higher reproductive fitness from avoiding inbreeding) and costs (mortality due to dispersal and disadvantage in competing for breeding opportunities) of dispersal, the evolutionarily stable strategy (ESS) of dispersal takes a very simple form for both males and females: $c_M = i\, k_F$ and $c_F = i\, k_M$. The elegant expressions can be understood intuitively as the dispersal cost counterbalancing the net cost of inbreeding depression. But if one takes this to predict that the system stabilizes at the internal equilibrium $k_M\, c_M = k_F\, c_F$, where the more-dispersive sex also incurs a higher dispersal cost, one would soon encounter a surprise. The internal equilibrium is unstable: if, for example, females are perturbed (e.g. by genetic drift) to disperse a little bit more, the selection on males to disperse will weaken, as the 'problem' of inbreeding is now reduced. Perturbations will eventually push the system to one of the boundary equilibria where only one sex disperses, and the other sex becomes completely philopatric.

Thus, even though inbreeding tolerance is predicted to differ between the sexes (Parker, 1979; Kokko & Ots, 2006), this difference does not yield solid predictions about which sex becomes dispersive and which sex becomes philopatric. The statement is typical of situations with positive feedback, where a system can reach two alternative evolutionarily stable states depending on initial conditions (Lehtonen & Kokko, 2012). Similar results where inbreeding leads to random initial conditions determining which sex ends up being the dispersive one are also shown in Gandon (1999), Perrin & Mazalov (2000), and Perrin & Goudet (2001).

Intriguingly, inbreeding tolerance also experiences another positive feedback: if inbreeding in a population as a whole is rare, we expect there to be many deleterious recessives and inbreeding depression will be strong (and inbreeding as a mating strategy is selected against); but if inbreeding occurs regularly, these recessives are exposed to selection and purged, and the system can thereafter move to tolerating inbreeding (Lande & Schemske, 1985). There is little work considering the interaction between the positive feedback involved in sex-specific dispersal and the other (possibly slower) feedback that is active in determining the strength of inbreeding depression. We are not aware of any analytical work in this area. Some authors have used simulations to examine the coevolution of inbreeding load and dispersal, the first one being Guillaume & Perrin (2006); under most (which they argue to be realistically mild) settings of deleterious mutations with a genomic mutation rate $U = 0.03$, their model did not produce significant differences between male and female dispersal. A bistable outcome where only one of the sexes disperses only emerged when parameter values were beyond what they consider a realistic range ($U = 1$). Note that their model considered minimal sexual asymmetry in reproductive strategy, implying that male and female variances in offspring number remain similar. It would be interesting to relax this assumption as well as combine it with results of Roze & Rousset (2009), whose model suggests that heterosis favours increased rates of dispersal, as dispersing individuals are unlikely to mate with patch mates who share the same deleterious recessives – but the genetic structure implemented in Roze & Rousset (2009) did not allow dispersal to be sex specific.

Why is it so difficult for models to produce consistently female-biased dispersal if the production of inbred young is more detrimental for female reproductive success? The key is to understand how dispersal incurs opportunity costs – i.e. removes chance of mating with certain individuals – in a manner that deviates from the set of assumptions used in classic models of inbreeding avoidance and tolerance (e.g. Parker, 1979). Assume, for a moment, that inbreeding avoidance is the sole reason to emigrate. Also assume that there is no mate choice (which might elevate the success of immigrant males if inbreeding is to be avoided) and, to keep the example as simple as possible, the mating success of any individual does not depend on its location *via* any other causality either. The logic is perhaps easiest to walk through for a 'winner takes all' situation where one of the males in each patch is randomly picked to fertilize all local females. If this male resides in his natal patch, and females are philopatric as well, then swapping some of these opportunities for unrelated females would improve his direct as well as indirect fitness (here we compute the direct fitness component only; the indirect component behaves analogously). Consider an example patch with four philopatric individuals (two males, two females) and four unrelated immigrants (again two males, two females). We assume that inbreeding halves offspring survival rate from $S$ to $S/2$, and a female produces only one offspring, but note that the logic works the same way for any level of female fecundity and inbreeding depression. If the female mates randomly in her natal patch, her expected number of offspring is ¼ $S/2$ + ¼ $S/2$ + ¼ $S$ + ¼ $S$ = ¾ $S$. If she disperses successfully and outbreeds, her expected number of offspring is $S$. For a male who fertilizes all the four females in a patch, the expected offspring production is 3$S$ if he stays at home, and 4$S$ if he has dispersed. For both sexes, dispersal boosts direct fitness by a factor of 4/3. The final step is to take into account that a male cannot predict whether he is going to be the local winner or not; the absolute value of expected fitness, and not just the fitness ratio, then becomes identical across the two sexes (in each deme the male wins with probability 1/4).

The above calculation seems to contradict classic results such as Parker (1979), where inbreeding tolerance was predicted to be higher for males than for females. The apparent contradiction arises because classic ways to phrase the problem ask whether a current mating opportunity should be rejected. For males in such a setting, accepting *versus* rejecting this particular opportunity might have negligible impact on his other opportunities. This way of thinking about the problem ceases to be correct when the trait under consideration is dispersal, as the opportunity costs brought about by moving are unavoidable. Dispersal causes some current potential mates to become unavailable (too far away) for any subsequent encounters with them, while others (likely unrelated potential mates) will be encountered. If the new encounters are better (less related) than the current ones, this is an equal boost to the direct fitness prospects of either sex. Indirect fitness calculations have exactly the same structure.

The toy example above, like most models of sex-biased dispersal, assumed that the choice of the sire is random. It is clear that expected fitness values can change if mating is not random, e.g. if females and/or males reject mating opportunities with kin. Here, a sexual asymmetry can become established more easily, with females choosier than males, because of asymmetry in the costs of choice becomes re-established: a female rejecting a mating usually does not mean that her eggs remain unfertilized (as long as she accepts some other local matings), while a male rejecting a mating represents a breeding opportunity that he truly loses. Lehmann & Perrin (2003) use mate choice by females to provide an interesting alternative explanation for the paradox of female-biased inbreeding cost and male-biased dispersal. The authors pointed out that females do not need to disperse to avoid inbreeding if they can recognize and reject related males as mates. This, in turn, creates selection for male dispersal because males suffer low mating success in their natal patch.

Lehmann & Perrin (2003) also point out that although inbreeding has a cost on fecundity and thus should be avoided, it also increases the inclusive fitness of an individual through the fecundity gains of its related mating partner. Therefore, at high inbreeding cost, females should reject kin matings and thereby cause male-biased dispersal, but at low inbreeding cost, inclusive fitness benefits should induce females to prefer related males, thereby promoting male philopatry (Lehmann & Perrin, 2003). The situation has further twists if cooperation among kin can also influence sex-biased dispersal through the avoidance or tolerance of inbreeding. If one sex benefits more from kin cooperation than the other sex, it can be selected to become more philopatric, and the feedback will then make the other sex disperse more to avoid inbreeding (Perrin & Goudet, 2001; Perrin & Lehmann, 2001).

The above causalities show how many factors can interact to produce the net outcome. Although difficult, it is valuable to tease out the effect of each single factor as much as possible. Perrin & Goudet (2001) provide us with a beautiful example. The authors study the joint effect of inbreeding avoidance and kin competition on the evolution of sex-biased dispersal. They first study a model that includes only local competition but has no inbreeding load, and then a model that includes inbreeding but excludes kin competition. After showing the effect of the two factors separately, the authors built a third model that includes the joint effect of both, and calculated the evolutionarily stable dispersal probability for each sex. Their results show that the effects of kin competition and inbreeding avoidance are not simply additive, because dispersal, even if it is induced by competition, prevents inbreeding, and makes the 'extra' dispersal largely pointless. In a fourth model, they then incorporated social interactions, with effects as explained above.

To summarize, the detrimental effects of inbreeding can drive sex-biased dispersal. Similar to kin competition, this mechanism works only when dispersal happens before mating, and the effect is strongest when deme size is small (high degrees of coancestry within the natal patch). But inbreeding differs from kin competition in that, to 'solve the problem' of inbreeding for both sexes, it is enough that only one of the sexes disperses. Models considering the effect of inbreeding avoidance alone often predict either no bias, or a bistable scenario (only one sex disperses while the other sex evolves to complete philopatry), or female-biased dispersal under polygyny, contradicting the general patterns observed in natural systems. But if females can recognize and refuse inbred matings, male-biased dispersal can evolve, as males maximize their chances when interacting with females outside the male's natal patch.

**(4) Fitness variance and effects of genetic architecture**

Above, we have mainly focused on theory that is based on expected (inclusive) fitness effects of dispersing, but there are at least two reasons why mean fitness of entire organisms is not the end of the story. First, fitness variance may matter for the evolution of traits, and the context of dispersal is a particularly important one, because by its very nature dispersal can impact how 'coarse-grained' the environment is that an evolving lineage encounters [it can qualify as a bet-hedging trait (Starrfelt & Kokko, 2012*a*) if it is costly for the individual in terms of mean fitness but also reduces fitness variance, e.g. by making lineages avoid extinction in ephemeral habitats]. Second, to the extent that dispersal control is genetic, it may also be based on different types of genetic architecture (Saastamoinen *et al.*, 2018) and this can have impacts on its sex specificity too.

While spatiotemporal environmental variation is generally a well-integrated part of the theoretical dispersal literature (e.g. Massol & Débarre, 2015), models commenting on sex biases are rare. Guillaume & Perrin (2009) discuss and model the intriguing sexual asymmetry that arises when inbreeding combines with polygyny. Under random mating, the fitness of a philopatric female varies more drastically than that of a dipersing female, because the former sometimes mates with related and sometimes with unrelated males, while the chance of a disperser mating with a relative is negligible. The argument for why the same asymmetry does not apply for males is clearest in the case of 'winner takes all', an extreme case of polygyny where one male mates with all the local females. While the fitness expectation of a male will differ between the natal patch and elsewhere (as inbreeding only occurs at the natal site; this is equally true for males as for females), a male who mates with everyone experiences much less fitness variation than a female does. Since selection can favour strategies that reduce fitness variance [bet-hedging, see Starrfelt & Kokko (2012*a*) for a review], there is more scope for females than males to benefit from reduced variance by dispersing. This is yet another nail in the coffin for any belief that polygyny should self-evidently predict male-biased dispersal.

Why, then, does polygyny often feature male-biased dispersal? By incorporating environmental and demographic stochasticity, the studies of Gros, Poethke & Hovestadt (2009) and Henry *et al.* (2016) shed new light on why males may evolve to be the more dispersive sex; in the case of Henry *et al.* (2016), this combines with a consideration of inbreeding avoidance. Stochasticity can promote male-biased dispersal under polygyny, if it helps to decouple the strength of local mate competition and local resource competition. Once male density is allowed to vary between patches, a randomly chosen male is more likely to be in a 'male-dense' patch than in a patch with few males; the benefit of moving (to a patch that potentially offers lighter competition) can therefore exceed the cost – with the net effect being more likely if demographic and/or environmental stochasticity are strong. Importantly, the effect is not self-evident, as female success may show spatiotemporal variation as well (and if females move, males may benefit from moving as they are otherwise left behind in patches with very few females; Meier *et al.*, 2011). For male-biased dispersal to be realized, the between-patch variance in success has to be greater for males than for females, which importantly differs from a mere within-patch expectation that arises very easily through sexual selection (e.g. one male siring all offspring in a patch – which does not on its own suffice to make males disperse more than females).

Turning to genetic architecture, Brom, Massot & Laloi (2018) recently re-examined a hypothesis (Whitney, 1976) that genetic sex determination might cause sex-biased dispersal if dispersal-related loci reside on the sex chromosomes, creating sex-linked inheritance and influencing relatedness patterns, as well as experiencing numerical asymmetries: in a population with a 1:1 sex ratio, there will be three times as many X (or Z) chromosomes as there are Y (or W) chromosomes. This leads to an exception of higher dispersal of the heterogametic sex (males in mammals, females in birds), but this prediction – while it fits the general 'Greenwoodian' pattern – is at its clearest under monogamy; the pattern becomes more complex when multiple mating is possible, because factors discussed above (e.g. between-patch variances in fitness for males and females) can also influence the evolving dispersal rates.

Finally, consider a special integrative case, that combines haplodiploidy with a metapopulation structure. Most of the literature operates under the (usually tacit) assumption that the species in question is diploid [see Saastamoinen *et al.* (2018) for a general discussion of genetic architecture assumptions in dispersal models]. If the relevant relatedness calculations differ between males and females, e.g. because of haplodiploidy, there are obvious new routes for how kin competition can induce sex-biased dispersal (Taylor, 1988). Females of haplodiploid species in the same habitat are often more related to each other than to the males, or the males to each other.

Under special conditions, asymmetries can also become important in diploid species, for example in systems where a proportion of individuals is infected by male-killing endosymbionts (Hurst & Jiggins, 2000). Once a female is infected, she can only produce infected daughters, whereas all male offspring are killed. Therefore, as modelled by Bonte, Hovestadt & Poethke (2009), relatedness becomes higher among males than among females in a partly infected habitat (there are only few males being produced, and these are more likely to share a mother compared with two randomly chosen local female offspring). The elevated relatedness among males leads to higher kin competition, which in turn induces female philopatry and male-biased dispersal.

Interestingly, the consequences of such male-biased dispersal negatively feed back to its underlying cause, especially under conditions of high dispersal cost and low environmental stochasticity. Because of the lack of males, infected local populations tend to go extinct, and are recolonized by uninfected populations. The coevolution of sex ratio, kin competition and sex-biased dispersal provides a fortuitous way for a metapopulation to 'cure' itself, escaping the spread of parasites. Local extinctions help avoid global extinction in this case. Such relatedness-asymmetry-induced dispersal bias in diploid populations might seem surprising, but in principle, any process that 'clumps' individuals into kin groups in one sex more than the other could cause selection to 'de-clump' them, assuming that kin competition plays a role and can be alleviated by dispersal. Although exciting in its dynamic richness, the model of Bonte *et al.* (2009) can thus be seen to be a complex example of a more general, simple, principle: if one sex currently disperses little and therefore 'clumps' and competes a lot with same-sex conspecifics, it is selected to disperse more (Perrin & Goudet, 2001). This feedback makes it perhaps understandable why we generally speak of dispersal biases – both sexes disperse, one more than the other – rather than completely unisexual dispersal patterns where one sex is the sole dispersal specialist.

To summarize, sex differences in the fitness variance across patches can drive sex-biased dispersal, with the general pattern that – all else being equal – the sex with larger between-patch fitness variance evolves to disperse more, although with dynamic feedback operating *via* local sex ratio fluctuations. Sex-specific genetic architecture can also be important for deriving correct predictions of sex-biased dispersal, e.g. genetic sex determination can matter if dispersal-controlling loci reside on sex chromosomes, and male-killing endosymbionts or haplodiploidy can interact with relatedness structure and local extinction–colonization dynamics, all impacting selection on sex-biased dispersal.

## III. LESSONS FOR EMPIRICISTS – AND FOR THEORETICIANS WRITING FOR THEM

Our main focus has been to highlight understudied gaps in theoretical work, but one can also ask how an empricist should react to the existing messages as well as the known gaps. We believe it would be counterproductive at this stage to produce a list of straightforward (let alone unidirectional) predictions, as the danger is that new empirical tests might be conducted without the hard work that comes with understanding the underlying theoretical rationale. We therefore instead highlight some of the issues our review has identified.

First, 'Greenwoodian' argumentation, where direct fitness effects (which sex suffers more from having to move) are given a major role, is very popular. The importance of familiarity with the natal habitat as a determinant of chances to acquire a breeding position is, however, often presumed, rather than directly tested; note that familiarity can also be gained over time through prospecting behaviour, and this does not have to happen at the natal site (Delgado *et al.*, 2014). We are not aware of direct attempts to estimate the relative competitiveness of immigrants *versus* philopatric individuals (i.e. $a_M$ and $a_F$ in the models of Perrin & Mazalov, 1999), which probably requires standardizing across body conditions as these can covary with the decision to stay *versus* disperse (or the ability to resist eviction). Also, sex-biased dispersal is surprisingly poorly linked to the extensive literature of prior residence effects [Stake, 2004; Strassmann & Queller, 2014; see Table A1 in Kokko, López-Sepulcre & Morrell (2006) for empirical examples], even though the connection is clear: emigrating means losing priority access to a resource, if this effect exists. Tests, of course, are difficult to conduct if one sex routinely disperses, or if individuals are not permitted to stay at home but are expelled – although removal experiments of competitors or parents can then shed light on the relative success.

Second, kin-based social interactions and inbreeding avoidance (*versus* tolerance) are popular empirical-study topics, but they have not developed their links to sex-biased dispersal theory, despite these aspects being well represented in this theory, as we have shown above [a disclaimer however is in place – our theoretically oriented scope does not allow us to cover existing empirical work in any great detail; we refer readers to Mabry *et al*. (2013) and Trochet *et al*. (2016)]. We suspect that the lack of contact between subfields is partly due to the challenges posed by the 'ghost of competition past': dispersal may evolve to alleviate competition between relatives to such a degree that competition appears not to be a problem that an organism has to deal with (see Section II.2*a*). This is not only a conceptual challenge; it has direct implications for what can be achieved in empirical enquiry. Dispersal inevitably prevents emigrants from interacting with their philopatric kin, and thus some of the questions are of a 'what if' nature. As an example: how strong would the (perhaps negative) fitness effect have been, if the individual in question had stayed and continued interacting (and mating) with the locals? This may be a less-appealing empirical research line than observing current interactions that do take place. Still, 'what if' questions are the bread and butter of experimental manipulations. Such manipulations and/or studies of populations under recent habitat fragmentation might help address what would happen to fitness if dispersal were restricted to a smaller scale than it currently operates over. One can hope to see more work conducted on quantifying competition for resources and mates, not only with respect to expected fitness but also its variance, and not forgetting the possibility of intersexual competition for resources of interest to both sexes simultaneously.

Finally, what should theoreticians do, when writing in a manner that empiricists (and thus science as a whole) will benefit from? While an empiricist might wish for a straightforward table with 'assumptions on the left and predicted patterns on the right', many predictions are not unidirectional, and as our review shows, the outcomes often feature a mix of several effects. While it is easy for a theoretician to 'switch off' mechanisms from operating (see Section II.2 for examples of how to block kin selection), experimental work has additionally to distinguish between plastic responses to experimental settings (e.g. Mishra *et al*., 2018; Van Petegem *et al*., 2018) and long-term evolutionary outcomes, including the potential for phylogenetic inertia (Mabry *et al*., 2013; Trochet *et al*., 2016). Theoretical papers become more helpful if they clearly state the expected patterns that would validate – or falsify – the ideas presented, and communication would improve if the idealized model outcomes were discussed against messy features of real life. For example, some models predict complete philopatry by one sex if inbreeding avoidance is the sole reason to disperse, but we also know that the long-term prospects of a population where females never disperse are poor; this does not make a model worthless, but its interpretation requires an understanding of how the philopatry prediction should be interpreted in real life. A proper dialogue between theory and empirical work will, ideally, combine different expertise angles to derive expectations when real systems incorporate multiple drivers – without falling into the opposite pit of saying 'everything interacts with everything, so we give up trying to make predictions'.

## IV. CONCLUSIONS

(1) The causal link between the mating system and sex-biased dispersal is not straightforward. A parallel work to ours (Trochet *et al.*, 2016) documenting empirical patterns (as opposed to theoretical expectations), also highlighted this point. Importantly, the fact that within a patch male mating success can vary more than female reproductive success is not sufficient to make males disperse more. Between-patch stochasticity, on the other hand, as well as more 'flexible' spatial arrangements (e.g. mating *en route*) can provide more robust causalities towards male-biased dispersal.

(2) The causalities that Greenwood's (1980) classic study considered, and the large body of theoretical work on sex-biased dispersal, are largely treated separately in the literature. The sex-specific home advantage of individuals (the ability to succeed in natal habitat *versus* elsewhere) and the possibility for them to evolve, are rarely included in models. Although the results might appear *a priori* too obvious (the sex with a sufficiently greater home advantage is selected to be philopatric under a wider range of conditions), we believe that new modelling exercises will be valuable, if only to re-establish how difficult it is to find consistent predictions linking polygyny to male-biased dispersal. Greenwood's writing also includes a relatively neglected part, where mate-searching by males extends to males dispersing more. This might be an easier way to associate polygyny with male-biased dispersal, but models with strict deme structure will miss this effect. The corresponding empirical challenge is to realize that dispersal may have evolved to make demes appear large and fluid enough that kin do not compete strongly at present, yet this does not mean kin competition did not drive the evolution of sex-biased dispersal — in a manner akin to the 'ghost of competition past'.

(3) While models have commented on the order of life-cycle events (which can be used to modify whether the presence of a related individual matters in the context of population regulation), they rarely consider that both males and females might partially depend on the same resources, even though the feeding niches of males and females typically overlap greatly, and the resource intake rates of males often exceeds that of females, at least in cases of male-biased sexual size dimorphism. Especially in iteroparous species (only rarely modelled), both male and female fitness will depend on current reproduction and also survival, thus emigration decision can have an impact on kin interactions not only within a sex, but also (to a great degree) across the sexes.

(4) There are surprising gaps in the literature where researchers appear to have jumped to study an exciting coevolutionary process (i.e. habitat-specific sex-ratio evolution interacting with competition that occurs either within a sex or jointly between the sexes) without looking explicitly at the simpler step that lacks adaptive sex-ratio responses. This does not make such work less valuable – it simply highlights that there is still space to work out the basics, hopefully with links to predictions that can be tested with ease in nature, such as the degree to which a male's presence harms or helps a local female's reproductive success or survival.

(5) Given that dispersal is a complex trait with aspects including the rate, distance distribution, and timing, models understandably vary in relevant assumptions and emphases, bringing along associated upsides and downsides. It is always advisable to be conscious of the likely effects that the modelling choices have on the findings. We hope our review will help to form an overview of the probable effects, e.g. why the timing of dispersal relative to mating is so often of crucial importance (Hirota, 2004, 2005; Wild & Taylor, 2004; Shaw & Kokko, 2014; Henry *et al.*, 2016).

(6) Our chosen focus on exposing the mathematical logic of the evolutionary causes of sex-biased dispersal has made us not only leave the relevant empirical evidence to outside the scope of this review [we direct readers to Trochet *et al.* (2016) for a recent review], but also omit many interesting models that mainly study the consequences, rather than causes, of sex-biased dispersal. Sex differences in dispersal can impact sex-ratio evolution, the evolution of social behaviours such as helping and harming, and the evolution of adaptive parental effects. Sex biases in dispersal rates or distances clearly also impact invasion and conservation biology. We hope that filling the gaps that currently exist in understanding the causes behind varying dispersal patterns will also help us to understand the consequences better.

# V. ACKNOWLEDGEMENTS

X.-Y.L. appreciates helpful discussions with Jean-Michel Guillon on a previous version of this review. Both authors would like to thank the two anonymous reviewers for their valuable comments, and the Swiss National Science Foundation for financial support.